\documentclass[conference]{IEEEtran}
\IEEEoverridecommandlockouts
\usepackage{cite}
\usepackage{amsmath,amssymb,amsfonts}
\usepackage{algorithmic}
\usepackage{graphicx}
\usepackage{textcomp}
\usepackage{xcolor}
\usepackage{booktabs}
\usepackage{multirow}
\usepackage{pgfplots}
\usepackage{lipsum}
\usepackage{soul}
\usepackage{url}
\pgfplotsset{compat=1.18}

\def\BibTeX{{\rm B\kern-.05em{\sc i\kern-.025em b}\kern-.08em
    T\kern-.1667em\lower.7ex\hbox{E}\kern-.125emX}}

\newcommand{\mat}[1]{\ensuremath{\operatorname{\boldsymbol{\MakeUppercase{#1}}}}} 

\newcommand{\R}{\ensuremath{\mathbb{R}}}    
\newcommand{\inRe}[1]{\in\R^{#1}} 
\newcommand{\tens}[1]{\ensuremath{\operatorname{\mathcal{#1}}}} 
\newcommand{\vect}[1]{\boldsymbol{#1}} 

\renewcommand{\O}[1]{\mathcal{O}\left( #1 \right)}
\newcommand{\out}{\otimes}
\newcommand{\khatrao}{\ensuremath{\odot}}
\newcommand{\hadamard}{\ensuremath{\circledast}}

\begin{document}

\title{Efficient Patient Fine-Tuned Seizure Detection with a Tensor Kernel Machine
}

\author{\IEEEauthorblockN{Seline J.S. de Rooij}
\IEEEauthorblockA{\textit{Signal Processing Systems} \\
\textit{Delft University of Technology}\\
Delft, Netherlands \\
{s.j.s.derooij@tudelft.nl}}
\and
\IEEEauthorblockN{Frederiek Wesel}
\IEEEauthorblockA{\textit{Delft Center for Systems and Control} \\
\textit{Delft University of Technology}\\
Delft, Netherlands \\
f.wesel@tudelft.nl}
\and
\IEEEauthorblockN{Borb\'ala Hunyadi}
\IEEEauthorblockA{\textit{Signal Processing Systems} \\
\textit{Delft University of Technology}\\
Delft, Netherlands \\
b.hunyadi@tudelft.nl}
}

\maketitle

\begin{abstract}
Recent developments in wearable devices have made accurate and efficient seizure detection more important than ever. A challenge in seizure detection is that patient-specific models typically outperform patient-independent models. However, in a wearable device one typically starts with a patient-independent model, until such patient-specific data is available. To avoid having to construct a new classifier with this data, as required in conventional kernel machines, we propose a transfer learning approach with a tensor kernel machine. This method learns the primal weights in a compressed form using the canonical polyadic decomposition, making it possible to efficiently update the weights of the patient-independent model with patient-specific data.
The results show that this patient fine-tuned model reaches as high a performance as a patient-specific SVM model with a model size that is twice as small as the patient-specific model and ten times as small as the patient-independent model. 
\end{abstract}

\begin{IEEEkeywords}
seizure detection, tensors, kernel machine, transfer learning
\end{IEEEkeywords}

\section{Introduction} \label{sec:intro}

\noindent Epilepsy is a common neurological disorder, it is characterized by abnormal electrical activity in the brain which causes seizures. Anti-epileptic drugs are only effective for about 70\% of the patients. Thus, the remaining 30\% continue to suffer from seizures, which impacts their quality of life \cite{WHO2019EpilepsyPublicHealth}. 

To improve this quality of life, wearable seizure detection devices are being developed, such as subcutaneous EEG \cite{weisdorf2019UltralongtermSubcutaneousHome}, wrist-worn \cite{beniczky2013DetectionGeneralizedTonic} or behind-the-ear EEG devices \cite{vandecasteele2020VisualSeizureAnnotation,you2022SemisupervisedAutomaticSeizure}. 
These devices could lead to more accurate diary keeping  \cite{vandecasteele2020VisualSeizureAnnotation}, and they may serve as a warning system when implemented in real-time \cite{naganur2022AutomatedSeizureDetection}.

This all requires an accurate and efficient seizure detection algorithm. 
Due to the large variability between epilepsy patients, patient-specific (PS)  seizure detection algorithms outperform their patient-independent (PI) counterparts \cite{siddiqui2020ReviewEpilepticSeizure}. 
However, there is often a lack of labelled patient-specific data, which makes it hard to implement a PS model.
 Furthermore, recent research has shown that there is also an increasing intra-patient variability over time \cite{schroeder2020SeizurePathwaysChange}. Suggesting, that even a patient-specific detector would need adaptation after some time.

This is why \textit{transfer learning} has become increasingly popular for seizure detection \cite{cui2023TransferLearningBased, wan2021ReviewTransferLearning}. In transfer learning information from a source task is used to improve the performance of a target task. While there are many different types of transfer learning, in this paper we consider the case of \textit{inductive} transfer learning, where both the source and the target task contain labels \cite{cui2023TransferLearningBased,pan2010SurveyTransferLearning}. Specifically, we consider the case where a patient-independent model is learned (\textit{source}), which is later updated with a limited amount of patient-specific data (the \textit{target} or \textit{patient fine-tuned} model). 

This type of transfer learning has been considered previously for seizure detection on wearable devices, using neural networks (CNN \cite{pisano2020ConvolutionalNeuralNetwork}) or kernel machines (SVM \cite{decooman2020PersonalizingHeartRateBased}). 
However, both of these methods have their limitations. Deep neural networks, for instance, typically contain a large number of model parameters and require a correspondingly large amount of data for training. This makes training computationally expensive. Although this could be remedied by fine-tuning only a few layers of the network \cite{tajbakhsh2016ConvolutionalNeuralNetworks}, such an approach may lead to suboptimal results and the resulting model remains large in size. 

Kernel machines, on the other hand, are well suited to small and medium size problems \cite{suykens2002LeastSquaresSupport}. However, because they are typically trained in their dual form, they can become intractable for large-scale problems due to the construction of the kernel matrix and because the number of model parameters scales with the training data (the \textit{support vectors}) \cite{suykens2002LeastSquaresSupport}. Transfer learning with a kernel machine can be done by learning a \textit{delta} model, to learn the difference between source and target \cite{yang2007AdaptingSVMClassifiers}. This delta model is then added to the source model.  The downside of this approach is that the update can only be done once and both the source and delta model need to be kept in memory. This makes it less suitable for a wearable device.







In this paper, we propose an efficient transfer learning method based on tensor kernel ridge regression (T-KRR) \cite{wesel2021LargeScaleLearningFourier}. The advantage of the T-KRR is that it learns the kernel ridge regression problem in the primal form and uses the canonical polyadic decomposition (CPD) \cite{kruskal1977ThreewayArraysRank} to learn the weights in compressed form. The use of the primal form and the CPD makes it suitable for large-scale and high-dimensional data. Furthermore, it becomes easier to adapt the model to new data by simply updating (part of) the weights.
To our knowledge, this is the first time the T-KRR classifier has been used for seizure detection and transfer learning.
\section{Classification Method} \label{sec:method}

\subsection{Notation and Preliminaries} \label{sec:tens_intro}
\noindent
Throughout this paper, the following notation conventions are used. Vectors and matrices are denoted by boldface lowercase and uppercase letters, respectively. Tensors, which are multidimensional arrays \cite{kolda2009TensorDecompositionsApplications}, are denoted by calligraphic letters, e.g. $\tens{A}$.

The symbols $\out$, $\khatrao$ and $\hadamard$ are used to denote the tensor outer product, the Khatri-Rao product and the Hadamard product, respectively.
The Frobenius inner product between two tensors is defined as, \mbox{$\langle \tens{A}, \tens{B} \rangle_\mathrm{F}:=\mathrm{vec}(\tens{A})^T\mathrm{vec}(\tens{B})$}, where \mbox{$\mathrm{vec}(\tens{A})_i = a_{i_1 i_2 \ldots i_D}$} is the vectorization of \mbox{$\tens{A} \inRe{I_1 \times \cdots \times I_D}$} to $\vect{a}\inRe{I}$ where $I=\prod_{d=1}^D I_d$. A tensor $\tens{A} \inRe{I_1 \times I_2 \times \cdots \times I_D}$ is considered to be \textit{rank-one} if it can be written as the outer product of $D$ vectors, \mbox{$\tens{A} = \vect{a}^{(1)} \out \vect{a}^{(2)} \out \cdots \out \vect{a}^{(D)}$} \cite{kolda2009TensorDecompositionsApplications}.

As the number of elements in a tensor grows exponentially in the number of dimensions, low-rank tensor decompositions are used to represent the tensor with a reduced number of parameters \cite{cichocki2016TensorNetworksDimensionalitya}. 
In this paper, we make use of the Canonical Polyadic decomposition (CPD) \cite{kruskal1977ThreewayArraysRank, hitchcock1927ExpressionTensorPolyadic}. A rank-$R$ CPD decomposes a tensor $\tens{A}\inRe{I_1 \times I_2 \times \cdots \times I_D}$ into a sum of $R$ rank-one tensors,

\begin{equation}
\tens{A} :=  \sum_{r=1}^R \mu_r \ \vect{a}_r^{(1)} \out \vect{a}_r^{(2)} \out \cdots \out \vect{a}_r^{(D)}.\label{eq:cpd} 
\end{equation}
Here, the vectors $\vect{a}_r^{(d)}$ are normalized to length one and the scaling is absorbed by $\vect{\mu}\inRe{R}$. 
The CPD can also be expressed as the Khatri-Rao product of \textit{factor matrices}, $$\text{vec}(\tens{A}):= \left(\mat{A}^{(D)} \khatrao \mat{A}^{(D-1)} \khatrao\cdots \khatrao \mat{A}^{(1)} \right) \vect{\mu} \ .$$
The columns of these factor matrices contain the vectors of the rank-one components, i.e. \mbox{$\mat{A}^{(d)} := [ \vect{a}_1^{(d)} \ \vect{a}_2^{(d)} \ \cdots \ \vect{a}_R^{(d)} ] \inRe{I_d \times R}$}.

For a sufficiently low rank this CPD results in a significant reduction in storage complexity compared to the full tensor, $\O{R \sum_d I_d}$ instead of $\O{\prod_d I_d}$.

\subsection{Kernel Machines} \label{sec:krr}
\noindent
In binary classification one aims to find a function $f(\cdot)$ that maps the input data $\vect{x} \inRe{D}$ to the correct label  $y\in \{-1,+1\}$. 

Kernel machines typically assume the following model form,
\begin{equation}
    f(\vect{x}) = \vect{\phi}(\vect{x})^T \vect{w} = \langle \vect{\phi} (\vect{x}) , \vect{w} \rangle_\mathrm{F},
\end{equation} 
where $\vect{w} \inRe{M}$  are the model weights. And \mbox{$\vect{\phi}(\cdot): \R^D \rightarrow \R^{M}$} is a feature map, which maps the input data to a higher dimensional feature space, a kernel Hilbert space, to enable nonlinear classification. The decision boundary for classification can be obtained by applying the sign function to the model response \cite{suykens2002LeastSquaresSupport}.  

The model weights $\vect{w}$ can be learned by minimizing a convex and  symmetric regularized loss function. An SVM \cite{cortes1995SupportvectorNetworks}, for instance, uses a hinge loss function, whereas kernel ridge regression (KRR) (or the LS-SVM) \cite{suykens2002LeastSquaresSupport} uses the  squared loss with Frobenius norm regularization. This leads to the following minimization problem,
\begin{equation}
\min_{\vect{w}} \sum_{n=1}^N \left(\left\langle \vect{\phi}\left(\vect{x}_n\right), \vect{w}\right\rangle_{\mathrm{F}} - y_n\right)^2+\lambda \langle \vect{w}, \vect{w} \rangle_{\mathrm{F}},\label{eq:primal_krr}
\end{equation}
for $N$ training samples.

Due to the high dimensionality of the feature map, $\phi(\cdot)$, often the \textit{kernel trick} is applied, which results in the non-parametric dual formulation. Therein, explicit feature maps only manifest as inner products, thus they can be replaced by a kernel function $\kappa(\vect{x}, \vect{x}') :=\langle \phi(\vect{x}),\phi(\vect{x}') \rangle_{\mathrm{F}}$.
The downside of this dual formulation is that it requires the calculation (and the formal inversion \cite{suykens2002LeastSquaresSupport}) of the kernel matrix \mbox{$\mat{K}\inRe{N\times N}$}, $K_{i,j} = \kappa (\vect{x}_i, \vect{x}_j)$, which becomes intractable for large $N$. 

\subsection{Tensor Kernel Ridge Regression} \label{sec:t-krr}
\noindent
In tensor kernel ridge regression (T-KRR) \cite{wesel2021LargeScaleLearningFourier} the following  feature mappings are considered, 
\begin{equation}
    \mat{\Phi}(\vect{x}) = \vect{\phi}^{(1)}(x_1) \out \vect{\phi}^{(2)}(x_2) \out \cdots \out \vect{\phi}^{(D)}(x_D), \label{eq:z_x}
\end{equation}
where $\mat{\Phi}(\cdot): \R^D \rightarrow \R^{M_1} \out \R^{M_2} \out \cdots \out \R^{M_D}$ is a tensor product of $D$ vectors and creates a rank one tensor. Here, $\vect{\phi}^{(d)}(\cdot): \R \rightarrow \R^{M_d}$ is the (local) feature map applied to the $d$-th dimension of the input. 

Using \eqref{eq:z_x} as the feature map we can reformulate the primal form of kernel ridge regression \eqref{eq:primal_krr} as follows,
\begin{equation}
    \min_{\tiny{\tens{W}}} \ \sum_{n=1}^N\left(\left\langle \mat{\Phi}(\vect{x}_n),\tens{W}\right\rangle_{\mathrm{F}}-y_n\right)^2+\lambda\langle\tens{W}, \tens{W}\rangle_{\mathrm{F}},
    \label{eq:min_problem_t_krr}
\end{equation}
where the weight vector $\vect{w}$ has been reshaped into a {$D$-dimensional} weight tensor $\tens{W} \inRe{M_1 \times M_2 \times \cdots \times M_D}$ by means of the vectorization identity (Section \ref{sec:tens_intro}).

As the size of the weight tensor scales exponentially with the dimension of the input data $D$, \eqref{eq:min_problem_t_krr} becomes intractable for high-dimensional input data. This is why in \cite{wesel2021LargeScaleLearningFourier} the weight tensor is constrained to have a rank-$R$ CPD structure.
The CPD weight tensor can be determined using an alternating linear scheme (ALS)\footnote{ For a detailed derivation please refer to \cite{wesel2021LargeScaleLearningFourier} and its appendix.}. Starting from some initial value, $\{\mat{W}^{(d)}_0\}_{d=1}^D$, the ALS algorithm updates the factor matrices of the CPD one at a time by solving,
\begin{equation} 
    \begin{aligned}
    {\min_{\tiny{\mat{W}^{(d)}}} \ \sum_{n=1}^N\left(\left\langle \vect{g}^{(d)}(\vect{x}_n),\textrm{vec}(\mat{W}^{(d)})\right\rangle_{\mathrm{F}} -y_n\right)^2} \\ {+ \lambda\left\langle\textrm{vec}\left(\mat{W}^{(d)^T}\mat{W}^{(d)}\right), \textrm{vec}\left(\mat{H}^{(d)}\right)\right\rangle_{\mathrm{F}}},
    \end{aligned}
    \label{eq:step_als_krr}
\end{equation}
where, \mbox{ \small $\boldsymbol{g}^{(d)}(\boldsymbol{x}):=\boldsymbol{\phi}^{(d)} \out\left(\boldsymbol{\phi}^{(1)^{\mathrm{T}}} \boldsymbol{W}^{(1)^{\mathrm{T}}} \hadamard \cdots \hadamard \boldsymbol{\phi}^{(D)^{\mathrm{T}}} \boldsymbol{W}^{(D)^{\mathrm{T}}}\right)$} and \mbox{ $\boldsymbol{H}^{(d)}:=\boldsymbol{W}^{(1)^{\mathrm{T}}} \boldsymbol{W}^{(1)} \hadamard \cdots \hadamard \boldsymbol{W}^{(D)^{\mathrm{T}}} \boldsymbol{W}^{(D)}$}. 

As in \cite{wesel2021LargeScaleLearningFourier} we approximate the RBF kernel with Laplace basis functions \cite{solin2020HilbertSpaceMethods}. Thus, the feature map is defined by, 
\begin{equation}
    \small{\left(\vect{\phi}^{(d)}\left(x_d\right)\right)_{i_d}=\frac{1}{\sqrt{U_d}} p\left(\frac{\pi i_d}{2 U_d}\right) \sin \left(\frac{\pi i_d\left(x_d+U_d\right)}{2 U_d}\right)}, \label{eq:fourier_feat}
\end{equation}
for input data centered in a hyperbox, $x_d \in [-U_d, U_d]$, and where $i_d=1, \ldots, {M}_d $ and $p(\cdot)$ is the spectral density of the RBF kernel \cite{rasmussen2005GaussianProcessesMachine}. 

\subsection{Transfer Learning with T-KRR}\label{sec:transfer_learn}
\noindent As stated in the introduction, in transfer learning knowledge from a \textit{source} task is transferred to a related \textit{target} task in order to improve its performance \cite{cui2023TransferLearningBased, pan2010SurveyTransferLearning}. A common technique to perform inductive transfer learning is \textit{fine-tuning}. In fine-tuning pre-trained weights from the source task are (partially) updated using data from the target task \cite{tajbakhsh2016ConvolutionalNeuralNetworks}. 

For the T-KRR, fine-tuning can easily be accomplished by using the weights of the source task as the initial value for the ALS optimization of the target task, and then updating (some of the) factor matrices of the CPD weight tensor using the target data, i.e. \mbox{$\forall d: \mat{W}^{(d)}_{0,\ \text{target}} =\mat{W}^{(d)}_{\text{source}}$}.

In this paper, we consider the specific case where the weights are first trained patient-independently and then \textit{fine-tuned} with a small amount of patient-specific data. 




\section{Data Processing} \label{sec:seizure}
\subsection{Data}
\noindent In this paper, EEG data from the Temple University Seizure Corpus (TUSZ, v1.5.2) \cite{obeid2016TempleUniversityHospital} was used to train and test the models. This is the largest publicly available EEG seizure detection dataset. It contains a total of $3.9\times 10^6$ seconds of annotated EEG data. Each recording session contains a \texttt{.txt} file with information about the patient and the seizure morphology. The dataset contains both EEG recordings that use the average reference montage and ones that use the linked ears reference montage. In our case, only the latter was used. 

Furthermore, we wanted to create a scenario that most closely reflects that of a wearable behind-the-ear EEG device, such as the ones presented in \cite{ vandecasteele2020VisualSeizureAnnotation, you2022SemisupervisedAutomaticSeizure}. Therefore, we used the T1 and T2 channels \cite{obretenova2021AdditionAnteriorTemporal}, located at the temporal lobe close to the ear, as a surrogate for the behind-the-ear channels. Additionally, we selected only patients who had temporal lobe seizures to ensure that the seizures were detectable with these channels. 
The following procedure was used to select the relevant recordings:
\begin{enumerate}
    \item Perform a keyword search on the \texttt{.txt} files for the terms ``left temporal" and ``right temporal".
    \item Check if the recordings contain the T1 and T2 channels. 
    \item Count the number of seizures per patient with a minimal seizure duration of 10 s.
    \item Select patients with $\geq 5$ seizures and perform a manual check of the \texttt{.txt} files to confirm the patient has temporal lobe seizures. 
\end{enumerate}
This leads us to $6$ patients with a median seizure count of $9$ ($\pm 4$) seizures per patient. In total, these recordings contain $7.3\times 10^4$ seconds of background data and $2.7 \times 10^4$ seconds of seizure data.  

\subsection{Preprocessing and Feature Extraction}
\begin{table}[t]
    \centering
        \caption{Extracted features \cite{vandecasteele2020VisualSeizureAnnotation, greene2008ComparisonQuantitativeEEG}.}
    \begin{tabular}{ p{1.2cm} p{6.3cm}  }
      \toprule
     \multirow{4}{5em}{Time domain} & 1-3. Number of zero crossings, maxima and minima\\
     & 4. Skewness\\
     & 5. Kurtosis\\
     & 6. RMS amplitude\\
     & 7. Linelength\\
     \midrule
    \multirow{4}{5em}{Frequency domain} & 8. Total power\\
    & 9. Peak frequency\\
    & 10-14. Mean power in frequency bands: $\delta$ (1-3 Hz), $\theta$ (4-8 Hz), $\alpha$ (9-13 Hz), $\beta$ (14-20 Hz), HF (40-80 Hz). \\
    \midrule
    \multirow{2}{5em}{Entropy} & 15. Spectral entropy\\
    & 16. Sample entropy\\
    \bottomrule
    \end{tabular}
    \label{tab:features}
\end{table}

\noindent All recordings were resampled to a sampling frequency of 250 Hz. Then, a bandpass filter (0.1-50 Hz, 4th order Butterworth) and a notch filter (50 Hz) was applied to each channel to reduce the presence of artifacts.  

After filtering the EEG data was divided into $2$s windows. For the seizure data 50\% overlap between windows was used, while no overlap was used for the non-seizure windows. This was done to balance the data for training. 
For each window features were extracted. The extracted features per channel are presented in Table \ref{tab:features}. It should be noted that the high-frequency band features (HF) were extracted prior to applying the low-pass filter. The presented features are a selection of features used in \cite{vandecasteele2020VisualSeizureAnnotation} with the addition of the line length feature, which was shown to work well for seizure detection \cite{greene2008ComparisonQuantitativeEEG}. Since we only use two EEG channels (T1 and T2), this amounts to a total of 32 features. 

Given our dataset contains recordings of seizures originating in both the left and right temporal lobes, we sort the features to remove this spatial structure \cite{vandecasteele2020VisualSeizureAnnotation}. Furthermore, all features (Table \ref{tab:features}) were scaled to lie in a unit hyperbox (i.e. \textit{min-max} scaling was used).

\section{Results and Discussion}   \label{sec:results}
\noindent All models described in this section were implemented in Python \cite{python}. The code and the corresponding models (with the used hyperparameters) can be found on Github.\footnote{\url{https://github.com/sderooij/PF-TKRR-public/}} 
\subsection{Validation and Hyperparameter Tuning}
\noindent The patient-independent (PI) models were evaluated using leave-one-patient-out cross-validation. The hyperparameters were tuned by gridsearch on the training set, using 5-fold cross-validation and maximizing the area under the receiver operating characteristic curve (AUROC). 

For the patient fine-tuned (PF) models, data from only one seizure (with some background data) was used to fine-tune the corresponding PI model using the same hyperparameters. Thus, in this case, a leave-one-seizure-\textit{in} cross-validation scheme was used. This validation scheme was also used for the patient-specific models. 

To evaluate the performance of the classifier we used the following metrics: AUROC, area under the precision-recall curve (AUPRC), F1-score, sensitivity and precision. 
The decision boundaries were chosen to maximize the F1-score. The overlapped windows were removed from the test set before calculating the performance metrics. 

\subsection{Classification Performance}
\begin{figure}[t]
    \centering
    \scalebox{0.55}{\input{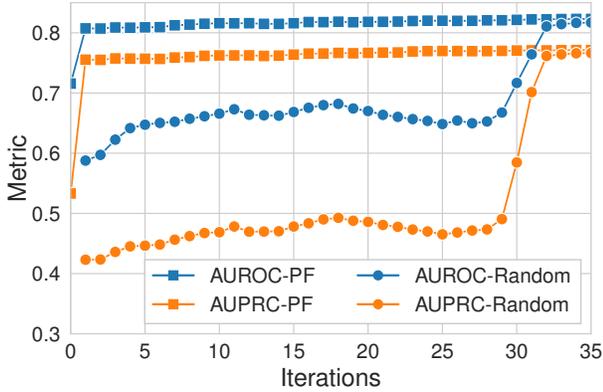}}
    \caption{AUC against the number of iterations of the ALS algorithm for T-KRR. Blue shows the AUROC and orange the AUPRC.
    Square dots represent the $\text{T-KRR}_{PF}$ model and round dots the $\text{T-KRR}_{PS}$ model with random initialization. }
    \label{fig:convergence}
\end{figure}

\noindent To investigate the influence of the number of iterations for fine-tuning the T-KRR, we plot the AUCPR and the AUROC (patient-mean) against the number of iterations (Figure \ref{fig:convergence}). An iteration equals solving \eqref{eq:step_als_krr}, i.e. updating one factor matrix of the weight tensor in CPD format. In the plot, we show two cases. One is the aforementioned fine-tuning approach, where the PI weights are used as the initial value for training with patient-specific data: the  $\text{T-KRR}_{PF}$ model. In the other case, the elements of the CPD weight tensor are initialized from a standard normal distribution and trained on the same PS data: the $\text{T-KRR}_{PS}$ model. 


From Figure \ref{fig:convergence} one can see that the $\text{T-KRR}_{PF}$ model has a significant increase in performance already after the first iteration compared to the patient-independent $\text{T-KRR}_{PI}$ model (at zero iterations). As the number of iterations grows, the performance increase is more gradual. The $\text{T-KRR}_{PS}$ model needs at least 32 iterations,  i.e. updating all the factor matrices, to reach the performance level of the $\text{T-KRR}_{PF}$ model.
Thus, starting from the PI weights yields a much faster convergence. 

Next, we compare the performance of the T-KRR classifier to that of a conventional SVM classifier with an RBF kernel. Both a patient-independent and a patient-specific SVM  model were constructed. 
For the $\text{T-KRR}_{PF}$ a full sweep of 32 iterations was used. 
Figure \ref{fig:barplot} provides barplot of the considered performance metrics. This barplot shows the mean performance across the patients and the corresponding 95\% confidence interval. 

In these results, we see that the performance of the $\text{T-KRR}_{PI}$ model seems slightly worse than the performance of the $\text{SVM}_{PI}$ model. This result may be due to the approximation of the RBF kernel \eqref{eq:fourier_feat} or the low-rank CPD approximation of the weight tensor. Furthermore, it is clear that being able to train on even a little bit of patient-specific data improves the performance significantly. The $\text{T-KRR}_{PF}$ model shows a comparable performance to the $\text{SVM}_{PS}$ model. 




\subsection{Model complexity} \label{}
\noindent Aside from the performance of the different classifiers, we also compare the complexity of the different models. When it comes to training an SVM, the cost is at least $\O{N^2}$ and at most $\O{N^3}$ \cite{chang2011LIBSVMLibrarySupport}, whereas for the T-KRR every ALS iteration costs $\O{ NM^2R^2}$, for $M=\max (M_d)$. 

\begin{figure}[t]
    \centering
    \scalebox{0.75}{\input{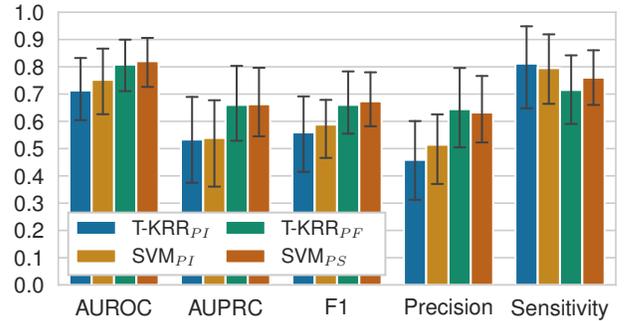}}
    \caption{Barplot of the performance for the different classifiers. The error bars show the 95\% confidence interval. }
    \label{fig:barplot}
\end{figure}

When it comes to inference, the complexity is dependent on the number of model parameters. For the SVM this is $\O{N_{SV}D}$, where $N_{SV}$ is the number of support vectors ($N_{SV}\leq N$). And for T-KRR this is $\O{DMR}$.
This means that for the models shown in Figure \ref{fig:barplot}, the $\textrm{T-KRR}_{PI}$ and $\textrm{T-KRR}_{PF}$ have the same number of parameters ($\sim 2 \times 10^4$), while the $\textrm{SVM}_{PI}$  models have a magnitude larger size than the $\textrm{SVM}_{PS}$ models  ($\sim 9 \times 10^5$ vs. $\sim 4 \times 10^4$ parameters).


\section{Conclusion} \label{sec:conclusion}
\noindent This paper investigated using the T-KRR classifier for seizure detection in a transfer learning setting. 
It was shown that this model is easily adaptable by using the previously trained weights as an initial value for training with new data. This concept was tested by updating the patient-independent model with a small amount of patient-specific data. The results showed that updating just one of the 32 factor matrices of the CPD weight tensor already improved the performance significantly. Ultimately, the patient fine-tuned model performed as well as a patient-specific SVM classifier trained on the same data, while using twice as few parameters.

Furthermore, for a patient-independent model, the performance of the T-KRR classifier is only slightly worse than that of the standardly used SVM classifier, while the T-KRR has ten times fewer parameters. This makes T-KRR classifier more suitable for a wearable device in case of strict memory requirements.

In future work, it would be of interest to investigate the effectiveness of the T-KRR classifier for \textit{online} transfer learning.
\section{Acknowledgements} \label{sec:acknowledge}
S.J.S. de Rooij and F. Wesel, and thereby this work, are supported by the TU Delft AI labs program. 

\bibliographystyle{IEEEtran}
\bibliography{IEEEabrv.bib, bibliography.bib}

\begin{thebibliography}{10}
\providecommand{\url}[1]{#1}
\csname url@samestyle\endcsname
\providecommand{\newblock}{\relax}
\providecommand{\bibinfo}[2]{#2}
\providecommand{\BIBentrySTDinterwordspacing}{\spaceskip=0pt\relax}
\providecommand{\BIBentryALTinterwordstretchfactor}{4}
\providecommand{\BIBentryALTinterwordspacing}{\spaceskip=\fontdimen2\font plus
\BIBentryALTinterwordstretchfactor\fontdimen3\font minus
  \fontdimen4\font\relax}
\providecommand{\BIBforeignlanguage}[2]{{%
\expandafter\ifx\csname l@#1\endcsname\relax
\typeout{** WARNING: IEEEtran.bst: No hyphenation pattern has been}%
\typeout{** loaded for the language `#1'. Using the pattern for}%
\typeout{** the default language instead.}%
\else
\language=\csname l@#1\endcsname
\fi
#2}}
\providecommand{\BIBdecl}{\relax}
\BIBdecl

\bibitem{WHO2019EpilepsyPublicHealth}
{World Health Organization}, ``{Epilepsy: a public health imperative:
  summary},'' Tech. Rep. WHO/MSD/MER/19.2, 2019.

\bibitem{weisdorf2019UltralongtermSubcutaneousHome}
S.~Weisdorf, J.~{Duun-Henriksen}, M.~J. Kjeldsen, F.~R. Poulsen, S.~W.
  Gangstad, and T.~W. Kj{\ae}r, ``Ultra-long-term subcutaneous home monitoring
  of epilepsy{\textemdash}490 days of {{EEG}} from nine patients,''
  \emph{Epilepsia}, vol.~60, no.~11, pp. 2204--2214, 2019.

\bibitem{beniczky2013DetectionGeneralizedTonic}
S.~Beniczky, T.~Polster, T.~W. Kjaer, and H.~Hjalgrim, ``Detection of
  generalized tonic{\textendash}clonic seizures by a wireless wrist
  accelerometer: {{A}} prospective, multicenter study,'' \emph{Epilepsia},
  vol.~54, no.~4, pp. e58--e61, 2013.

\bibitem{vandecasteele2020VisualSeizureAnnotation}
K.~Vandecasteele, T.~De~Cooman, J.~Dan, E.~Cleeren, S.~Van~Huffel, B.~Hunyadi,
  and W.~Van~Paesschen, ``Visual seizure annotation and automated seizure
  detection using behind-the-ear electroencephalographic channels,''
  \emph{Epilepsia}, vol.~61, no.~4, pp. 766--775, Apr. 2020.

\bibitem{you2022SemisupervisedAutomaticSeizure}
S.~You, B.~Hwan~Cho, Y.-M. Shon, D.-W. Seo, and I.~Y. Kim, ``Semi-supervised
  automatic seizure detection using personalized anomaly detecting variational
  autoencoder with behind-the-ear {{EEG}},'' \emph{Computer Methods and
  Programs in Biomedicine}, vol. 213, p. 106542, Jan. 2022.

\bibitem{naganur2022AutomatedSeizureDetection}
V.~Naganur, S.~Sivathamboo, Z.~Chen, S.~Kusmakar, A.~{Antonic-Baker}, T.~J.
  O'Brien, and P.~Kwan, ``Automated seizure detection with noninvasive wearable
  devices: {{A}} systematic review and meta-analysis,'' \emph{Epilepsia},
  vol.~63, no.~8, pp. 1930--1941, 2022.

\bibitem{siddiqui2020ReviewEpilepticSeizure}
M.~K. Siddiqui, R.~{Morales-Menendez}, X.~Huang, and N.~Hussain, ``A review of
  epileptic seizure detection using machine learning classifiers,'' \emph{Brain
  Informatics}, vol.~7, no.~1, p.~5, Dec. 2020.

\bibitem{schroeder2020SeizurePathwaysChange}
G.~M. Schroeder, B.~Diehl, F.~A. Chowdhury, J.~S. Duncan, J.~{de Tisi}, A.~J.
  Trevelyan, R.~Forsyth, A.~Jackson, P.~N. Taylor, and Y.~Wang, ``Seizure
  pathways change on circadian and slower timescales in individual patients
  with focal epilepsy,'' \emph{Proceedings of the National Academy of
  Sciences}, vol. 117, no.~20, pp. 11\,048--11\,058, May 2020.

\bibitem{cui2023TransferLearningBased}
X.~Cui, J.~Cao, T.~Jiang, and F.~Gao, ``Transfer {{Learning Based Seizure
  Detection}}: {{A Review}},'' in \emph{Cognitive {{Computation}} and
  {{Systems}}}, ser. Communications in {{Computer}} and {{Information
  Science}}, F.~Sun, J.~Li, H.~Liu, and Z.~Chu, Eds.\hskip 1em plus 0.5em minus
  0.4em\relax {Singapore}: {Springer Nature}, 2023, pp. 160--175.

\bibitem{wan2021ReviewTransferLearning}
Z.~Wan, R.~Yang, M.~Huang, N.~Zeng, and X.~Liu, ``A review on transfer learning
  in {{EEG}} signal analysis,'' \emph{Neurocomputing}, vol. 421, pp. 1--14,
  Jan. 2021.

\bibitem{pan2010SurveyTransferLearning}
S.~J. Pan and Q.~Yang, ``A {{Survey}} on {{Transfer Learning}},'' \emph{IEEE
  Transactions on Knowledge and Data Engineering}, vol.~22, no.~10, pp.
  1345--1359, Oct. 2010.

\bibitem{pisano2020ConvolutionalNeuralNetwork}
F.~Pisano, G.~Sias, A.~Fanni, B.~Cannas, A.~Dourado, B.~Pisano, and C.~A.
  Teixeira, ``Convolutional {{Neural Network}} for {{Seizure Detection}} of
  {{Nocturnal Frontal Lobe Epilepsy}},'' \emph{Complexity}, vol. 2020, p.
  e4825767, Mar. 2020.

\bibitem{decooman2020PersonalizingHeartRateBased}
T.~De~Cooman, K.~Vandecasteele, C.~Varon, B.~Hunyadi, E.~Cleeren,
  W.~Van~Paesschen, and S.~Van~Huffel, ``Personalizing {{Heart Rate-Based
  Seizure Detection Using Supervised SVM Transfer Learning}},'' \emph{Frontiers
  in Neurology}, vol.~11, 2020.

\bibitem{tajbakhsh2016ConvolutionalNeuralNetworks}
N.~Tajbakhsh, J.~Y. Shin, S.~R. Gurudu, R.~T. Hurst, C.~B. Kendall, M.~B.
  Gotway, and J.~Liang, ``Convolutional {{Neural Networks}} for {{Medical Image
  Analysis}}: {{Full Training}} or {{Fine Tuning}}?'' \emph{IEEE Transactions
  on Medical Imaging}, vol.~35, no.~5, pp. 1299--1312, May 2016.

\bibitem{suykens2002LeastSquaresSupport}
J.~A.~K. Suykens, T.~Van~Gestel, J.~De~Brabanter, B.~De~Moor, and
  J.~Vandewalle, \emph{Least {{Squares Support Vector Machines}}}.\hskip 1em
  plus 0.5em minus 0.4em\relax {World Scientific}, Nov. 2002.

\bibitem{yang2007AdaptingSVMClassifiers}
J.~Yang, R.~Yan, and A.~G. Hauptmann, ``Adapting {{SVM Classifiers}} to
  {{Data}} with {{Shifted Distributions}},'' in \emph{Seventh {{IEEE
  International Conference}} on {{Data Mining Workshops}} ({{ICDMW}} 2007)},
  Oct. 2007, pp. 69--76.

\bibitem{wesel2021LargeScaleLearningFourier}
F.~Wesel and K.~Batselier, ``Large-{{Scale Learning}} with {{Fourier Features}}
  and {{Tensor Decompositions}},'' in \emph{Advances in {{Neural Information
  Processing Systems}}}, vol.~34.\hskip 1em plus 0.5em minus 0.4em\relax
  {Curran Associates, Inc.}, 2021, pp. 17\,543--17\,554.

\bibitem{kruskal1977ThreewayArraysRank}
J.~B. Kruskal, ``Three-way arrays: Rank and uniqueness of trilinear
  decompositions, with application to arithmetic complexity and statistics,''
  \emph{Linear Algebra and its Applications}, vol.~18, no.~2, pp. 95--138, Jan.
  1977.

\bibitem{kolda2009TensorDecompositionsApplications}
T.~G. Kolda and B.~W. Bader, ``Tensor {{Decompositions}} and
  {{Applications}},'' \emph{SIAM Review}, vol.~51, no.~3, pp. 455--500, Aug.
  2009.

\bibitem{cichocki2016TensorNetworksDimensionalitya}
A.~Cichocki, N.~Lee, I.~Oseledets, A.-H. Phan, Q.~Zhao, and D.~P. Mandic,
  ``Tensor {{Networks}} for {{Dimensionality Reduction}} and {{Large-scale
  Optimization}}: {{Part}} 1 {{Low-Rank Tensor Decompositions}},''
  \emph{Foundations and Trends{\textregistered} in Machine Learning}, vol.~9,
  no. 4-5, pp. 249--429, 2016.

\bibitem{hitchcock1927ExpressionTensorPolyadic}
F.~L. Hitchcock, ``The {{Expression}} of a {{Tensor}} or a {{Polyadic}} as a
  {{Sum}} of {{Products}},'' \emph{Journal of Mathematics and Physics}, vol.~6,
  no. 1-4, pp. 164--189, 1927.

\bibitem{cortes1995SupportvectorNetworks}
C.~Cortes and V.~Vapnik, ``Support-vector networks,'' \emph{Machine Learning},
  vol.~20, no.~3, pp. 273--297, Sep. 1995.

\bibitem{solin2020HilbertSpaceMethods}
A.~Solin and S.~S{\"a}rkk{\"a}, ``Hilbert space methods for reduced-rank
  {{Gaussian}} process regression,'' \emph{Statistics and Computing}, vol.~30,
  no.~2, pp. 419--446, Mar. 2020.

\bibitem{rasmussen2005GaussianProcessesMachine}
C.~E. Rasmussen and C.~K.~I. Williams, \emph{Gaussian {{Processes}} for
  {{Machine Learning}}}.\hskip 1em plus 0.5em minus 0.4em\relax {The MIT
  Press}, Nov. 2005.

\bibitem{obeid2016TempleUniversityHospital}
I.~Obeid and J.~Picone, ``The {{Temple University Hospital EEG Data Corpus}},''
  \emph{Frontiers in Neuroscience}, vol.~10, May 2016.

\bibitem{obretenova2021AdditionAnteriorTemporal}
S.~Obretenova, M.~F. Villamar, and S.~Tobochnik, ``Addition of {{Anterior
  Temporal EEG Electrodes}} to {{Improve Seizure Detection}},'' \emph{The
  Neurohospitalist}, vol.~11, no.~1, pp. 89--90, Jan. 2021.

\bibitem{greene2008ComparisonQuantitativeEEG}
B.~R. Greene, S.~Faul, W.~P. Marnane, G.~Lightbody, I.~Korotchikova, and G.~B.
  Boylan, ``A comparison of quantitative {{EEG}} features for neonatal seizure
  detection,'' \emph{Clinical Neurophysiology}, vol. 119, no.~6, pp.
  1248--1261, Jun. 2008.

\bibitem{python}
G.~Van~Rossum and F.~L. Drake, \emph{Python 3 Reference Manual}.\hskip 1em plus
  0.5em minus 0.4em\relax {Scotts Valley, CA}: {CreateSpace}, 2009.

\bibitem{chang2011LIBSVMLibrarySupport}
C.-C. Chang and C.-J. Lin, ``{{LIBSVM}}: {{A}} library for support vector
  machines,'' \emph{ACM Transactions on Intelligent Systems and Technology},
  vol.~2, no.~3, pp. 27:1--27:27, May 2011.

\end{thebibliography}

\end{document}